\documentclass[%
longbibliography,
letterpaper,
reprint,
showpacs,
 amsmath,amssymb,
 aps,
]{revtex4-1}

\usepackage{multirow}
\usepackage{color}
\usepackage{capt-of}

\definecolor{red}{RGB}{255,0,0}
\definecolor{blue}{RGB}{0,0,255}
\definecolor{purple}{RGB}{128,0,128}
\definecolor{green}{RGB}{0,128,0}
\definecolor{brown}{RGB}{128,0,0}
\definecolor{pink}{RGB}{255,0,255}
\definecolor{cyan}{RGB}{0,255,255}
\definecolor{orange}{RGB}{255,128,0}

\usepackage{graphicx}% Include figure files
\usepackage{dcolumn}% Align table columns on decimal point
\usepackage{bm}% bold math
\begin{document}

\title{The interpretation of the field angle dependence of the critical current in defect-engineered superconductors}

\author{Stuart C.\ Wimbush}
 \email{s.wimbush@irl.cri.nz.}
\author{Nicholas J.\ Long}
 \email{n.long@irl.cri.nz.}
\affiliation{Industrial Research Limited, 69 Gracefield Road, Lower Hutt 5010, New Zealand.}

\date{\today}

\begin{abstract}
We apply the vortex path model of critical currents to a comprehensive analysis of contemporary data on defect-engineered superconductors, showing that it provides a consistent and detailed interpretation of the experimental data for a diverse range of materials. We address the question of whether electron mass anisotropy plays a role of any consequence in determining the form of this data and conclude that it does not. By abandoning this false interpretation of the data, we are able to make significant progress in understanding the real origin of the observed behavior. In particular, we are able to explain a number of common features in the data including shoulders at intermediate angles, a uniform response over a wide angular range and the greater discrimination between individual defect populations at higher fields. We also correct several misconceptions including the idea that a peak in the angular dependence of the critical current is a necessary signature of strong correlated pinning, and conversely that the existence of such a peak implies the existence of correlated pinning aligned to the particular direction. The consistency of the vortex path model with the principle of maximum entropy is introduced.
\end{abstract}

\maketitle

\section{\label{sec:intro}Introduction}

In the field of high temperature superconductor (HTS) research and development, a significant hurdle has proved to be the anisotropic properties of the available materials. For a typical YBa$_2$Cu$_3$O$_7$ (YBCO) sample, the variation in the critical current density with the angle of an applied magnetic field, $J_c(\theta)$, can exceed a factor ten at high fields \cite{Coulter1999}. Successful methods of tackling this issue have been developed, most notable of which has been the incorporation of artificial pinning centers in the form of self-assembled BaZrO$_3$ nanocolumns \cite{Driscoll2004}. These have introduced a so-called $c$-axis peak to the pinning profile, enhancing $J_c$ for fields applied in the $c$-direction, perpendicular to the film surface. Rather than solving the problem, however, this has merely shifted the region of concern to intermediate angles, between the two peaks, where $J_c$ remains low. Adding BaZrO$_3$ nanoparticles creates a more isotropic profile but the effect varies substantially with field and temperature \cite{Miura2011}. In order to fully overcome this issue by engineering appropriate defect structures, we need first to understand the origin of the $J_c(\theta)$ dependence. Remarkably, no consensus model of $J_c(\theta)$ exists, highlighting the general weakness in established models of $J_c$ to reproduce the phenomenology of critical currents.

One proposed origin of at least some of the features of $J_c(\theta)$ gaining momentum has been the electronic mass anisotropy of the materials \cite{Xu1994,Kumar1994,CivaleJLTP2004,*CivaleAPL2004}. However, the physical link between this intrinsic material anisotropy and (extrinsic) properties such as $J_c$ has not been made clear. While application of this approach to understand the physics of pristine samples of clean superconductors \cite{Kidszun2011} may have some merit, its direct transfer to explaining the pinning in technological materials incorporating artificial pinning centers, resulting in ``effective electronic mass anisotropy ratios'' that bear no relation to the fundamental material parameters \cite{Gutierrez2007} is questionable. We therefore begin by reviewing the physics of the electron mass anisotropy scaling approach and discuss whether this approach is applicable to high-$J_c$ pinning-engineered samples.

Additional to this, many recent pinning-optimized samples \cite{Maiorov2007} exhibit features such as pronounced shoulders \cite{Chudy2010} and flat angular regions \cite{Mikheenko2011} in $J_c(\theta)$ that cannot be explained by the naive summation of an $ab$-plane intrinsic pinning peak, a $c$-axis correlated pinning peak and a random pinning contribution characterized by a geometrical anisotropy of whatever origin. The presence of similar features throughout the literature from the earliest samples produced \cite{Roas1990,BraithwaiteJLTP1993a,*BraithwaiteJLTP1993b} to present-day `champion' samples \cite{Maiorov2009} means that these can not be dismissed as due to poor sample quality or inaccurate measurement, but must be acknowledged as genuine effects of an optimized pinning landscape. Combine this with the complex temperature and magnetic field dependence of $J_c$ now starting to be revealed \cite{Xu2010} and it is apparent that a deeper understanding of the effects of a complex flux pinning landscape is required.

Here we demonstrate that an analysis of the statistical effects of combining $c$-axis correlated pinning and $ab$-plane correlated pinning (whether intrinsic, extrinsic, or surface in origin) can explain all the features observed to date in pinning-optimized samples. Isotropic pinning resulting from point defects or incorporated second-phase nanoparticles contributes to this picture in the form of `bridges' between these defining pinning populations. The aim of this paper is to show that the model that results from this analysis, termed the vortex path model, provides a consistent and detailed explanation of the experimental features that have risen to prominence in recent studies. Using this model, $J_c(\theta)$ is decomposed into statistically distinguishable components that can be related back to distinct aspects of the sample microstructure that determines $J_c$ --- a link that is missing in other approaches.

\section{\label{sec:models}Models of $J_c(\theta)$}

\subsection{\label{sec:anisotropy}Electron mass anisotropy scaling}

The electron mass anisotropy scaling approach has recently become a common method of analyzing $J_c(\theta)$ data obtained from diverse species of superconducting materials \cite{Choi2004,Paturi2008,Iida2010,Pardo2011}. In some cases misinterpretation of the process has lead to simple fits of the Ginzburg-Landau mass anisotropy function to $J_c(\theta)$ being employed \cite{Crisan2010,*Dang2010}. It is therefore worth reviewing the reasoning behind this approach to understand the limits of its applicability. If the scaling approach is valid then this affects how we should interpret the various features of $J_c(\theta)$.

The scaling approach was originally proposed by Blatter \emph{et al.}\ \citetext{\citealp{[(a) ]Blatter1992,*[(b) ]Blatter1994,*[(c) ]Blatter2008}(a)}, with a similar formulation being independently proposed by Hao and Clem \cite{Hao1992}. The idea is to apply results derived for an isotropic superconductor to an anisotropic superconductor through a rescaling of the magnetic field, temperature, characteristic lengths, etc. The scaling rules are obtained by finding a transformation which recasts the Gibbs free energy of an anisotropic superconductor into the form for an isotropic material. The primary result is then a scaling rule such as
\begin{equation}
Q(\theta,H,T,\xi,\lambda_L,\varepsilon,\delta) = s_Q\tilde{Q}(\varepsilon_\theta H,T/\varepsilon,\xi,\lambda_L,\delta/\varepsilon).
\label{eq:scalingrule}
\end{equation}
Blatter \emph{et al.}\ then apply this rule to calculate critical currents for the case of single-vortex weak collective pinning, that is pinning by randomly distributed point pins. The results are $J_c^\parallel = J_c^c$, independent of $\theta$, for the in-plane critical current density (as usually measured in thin film experiments), where $J_c^c$ is the in-plane $J_c$ with the field applied in the $c$-direction, and $J_c^\perp = \varepsilon_\theta J_c^c$ for the out-of-plane critical current density, with $\varepsilon_\theta^2 = \varepsilon(\theta) = \cos^2\theta + \varepsilon^2\sin^2\theta$, where $\varepsilon^2 = m_{ab}/m_c$ is the electronic mass anisotropy. For consistency throughout our paper, and in contrast to Blatter, we have taken $\theta = 0$ to be perpendicular to the $ab$-planes.

The result $J_c^\parallel = \mathrm{constant}$ for low fields is consistently ignored by those who use the scaling approach to analyse $J_c(\theta)$ data. We are not aware of any report claiming to have confirmed this result experimentally, thereby validating the scaling hypothesis. Many samples reported in this paper and elsewhere have wide regions of $J_c^\parallel = \mathrm{constant}$ at low field. However, these samples also have correlated $c$-axis defects such as grain boundaries, threading dislocations and twin plane intersections and it would be puzzling to assume these made no contribution to pinning in these circumstances.

The scaling procedure that is widely used in the literature is based on an expression postulated in a subsequent review by Blatter \emph{et al.}\ \citetext{\citealp{Blatter1994}(b)} for the high field case, defined by $H(\theta) > \beta_{sb}(J_c^c/J_d)H_{c2}(\theta)$ where $\beta_{sb} \approx 5$ and $J_d$ is the depairing current density. In this case, no physical model of $J_c$ to which the scaling rules can directly be related is proposed. Rather, the authors simply assert that at high field, $J_c$ will scale in accordance with $H \rightarrow \varepsilon_\theta H$, i.e.
\begin{equation}
J_c^\parallel(\theta,H) = J_c^c(\varepsilon_\theta H).
\label{eq:massanisotropyscaling}
\end{equation}
To evaluate the status of this assertion we recall that the scaling rule, Eq.~(\ref{eq:scalingrule}), is a non-unique mathematical transformation that allows us to write the equilibrium free energy for an anisotropic superconductor in the same form as for an isotropic superconductor. It is not a physical theory where a \emph{measurement} of a physical quantity may actually give a different answer depending on the value of another physical variable. In particular, a \emph{measurement} of the field in the sample does not scale with these rules. Any change in the critical current must therefore have an indirect mechanism, which this approach does not address. The critical current density is a non-equilibrium property of the sample; formally, it is a parameter in a constitutive equation such as $E = E_0(J/J_c)^n$. There is no formal derivation possible from Eq.~(\ref{eq:scalingrule}) to Eq.~(\ref{eq:massanisotropyscaling}); rather, Eq.~(\ref{eq:massanisotropyscaling}) is a new hypothesis. The justification for the adoption of this hypothesis, noted in both \citetext{\citealp{Blatter1994}(b)} and \cite{Hao1992}, is a perceived similarity between the high field behavior of $J_c(\theta)$ and the field-scaling rule for the free energy. The justification given is empirical, not theoretical.

This justification, however, is not well founded. If a broad peak in $J_c(\theta)$ is determined by a variation in pinning density with angle, for example arising from the interaction of correlated defects, rather than from the mass anisotropy of the material, then we would still expect $J_c(\theta)$ data to be approximately scaleable in this way. Typically, for HTS samples, critical currents in large out-of-plane fields behave as $J_c(H{\parallel}c) \propto H^{-\alpha}$. If, for an in-plane field, we have a higher density of correlated pinning defects but otherwise similar physics we would expect $J_c(H{\parallel}ab) \propto (H/k)^{-\alpha}$, where $k$ is a measure of the change in pinning density between the field directions. We can scale $H \rightarrow H\varepsilon(k,\theta)$ so the data coincide at these two angles using an equation of the same form as the mass anisotropy, $\varepsilon^2(k,\theta) = \cos^2\theta + k^2\sin^2\theta$. Further, if $J_c(\theta)$ varies smoothly and broadly then $\varepsilon(k,\theta)$ will also give a reasonable fit to all the intermediate angle data. The electron mass anisotropy expression can therefore be a reasonable fitting function whether the origin of the behavior is in fact electronic mass anisotropy or simply pinning density anisotropy arising from correlated defects. Experimentalists who employ this approach assume that correlated pinning can only be effective and therefore influence their data over narrow angular ranges. The evidence from samples implanted at different angles by heavy ions \cite{Holzapfel1993, Strickland2009} or the addition of self-assembled nanocolumns \cite{Goyal2005} suggests this confidence is misplaced. Similar to the low field result, those who apply this scaling also consistently ignore the lower bound to the field range over which it is applicable and do not attempt to verify any crossover from the low field to the high field regime. The lower bound quantified above can be significant for HTS since $H_{c2}$ is large; at low temperature it may easily be tens of teslas, beyond most common experimental capability.

Aside from these criticisms the scaling approach suffers from one further fatal flaw when applied to pinning-optimized samples: it ignores the pinning summation problem. If we add random isotropic defects to a sample that already possesses a high density of $ab$-plane pinning but much lower $c$-axis pinning then we expect the effect on $J_c(\theta)$ about the $c$-axis direction to be much stronger than the effect about the $ab$-plane. This is exactly what is seen experimentally \cite{Miura2011,Chudy2010,Long2005}. As individual vortices interact with multiple defects, correlated and uncorrelated, the effect on $J_c$ of additions to the pinning landscape depends on the pinning already present and the scaling approach has no way of taking this into account.

The use of the high field scaling approach to analyze $J_c(\theta)$ has led to a situation where a correspondence between the experimental data and the known defect structures of HTS films is missing. We think this approach is wrong and should be abandoned. In the next section we recapitulate a model that can distinguish how different defect populations within the sample contribute to $J_c$, and that can enumerate how these distinct contributions combine to yield the overall $J_c(\theta)$.

\subsection{\label{sec:VPM}The vortex path model}

A statistical mechanical approach to modeling the response of a superconductor to a particular population of pinning defects was first proposed in \cite{Long2007} and formally presented in \cite{Long2008}. It was subsequently termed the vortex path model in \cite{Paturi2010} since at its core, it considers the multiplicity of possible pinned vortex paths through a sample, as determined by its defect structure, to quantify the relative pinning strength for fields applied at different angles. It has recently been used to describe experimental data where the electronic mass anisotropy approach has been found inadequate \cite{Mikheenko2010}.

Here we do not repeat the full exposition of the model but instead give a general theoretical context as an introduction. The model is constructed in the spirit of a maximum entropy approach \cite{JaynesPR1957a,*JaynesPR1957b}. If we consider any macroscopic property of a system, where the macroscopic state can be related to a multiplicity of microstates, then to predict the behavior of the property we maximize the Shannon entropy of the distribution of microstates of the system subject to the known constraints. For elementary constraints this leads to the ubiquitous functions of physics; for example, the Gaussian function is obtained upon maximizing the Shannon entropy subject to the constraints that the mean and the variance of the distribution of microstates are fixed, while the (truncated) Lorentzian function is obtained when only the variance is fixed \cite{Carazza1976}. The critical current density of a superconductor is just such a macroscopic property arising from the multiplicity of microstates in which vortices are pinned at a particular angle, and following \cite{Long2008} we can write
\begin{equation}
J_c(\theta) = J_0 g(\theta),
\label{eq:Jc}
\end{equation}
where $g(\theta)$ is the density distribution of these pinning microstates normalized such that ${\int_0^\pi}g(\theta)d\theta = 1$ and the proportionality factor $J_0 = {\int_0^\pi}J_c(\theta)d\theta$. In \cite{Long2008}, a Lorentz force definition of the critical current was used to define $J_0$ in terms of the pinning force. However, this definition of $J_c$ is unnecessary, and instead Eq.~\ref{eq:Jc} itself serves to define the critical current. We can then maximize the Shannon entropy
\begin{equation}
H = -{\int_0^\pi}g(\theta)\ln(g(\theta))d\theta
\label{eq:entropy}
\end{equation}
with appropriate constraints in order to predict $J_c(\theta)$. Without any constraints the result is a uniform distribution and we obtain a constant $J_c$,
\begin{subequations}
\label{eq:VPMequations}
\begin{equation}
J_c(\theta) = \frac{J_0}{\pi}.
\label{eq:uniform}
\end{equation}

For a sample containing correlated defects within the $ab$-plane and along the $c$-axis we know that $J_c(\theta)$ may exhibit peaks centered on the directions of the defects. These peaks are broadened by the vortices' interactions with the orthogonal defect populations. The vortex path model introduces constraints on $g(\theta)$ by requiring the vortex microstates to follow a path which interacts with these defects. In the usual geometry for $J_c$ measurements we define $\theta = \tan^{-1}(y/z)$, where $\hat{y}$ lies within the plane of the sample and $\hat{z}$ is perpendicular to it. For a peak centered on $\hat{y}$, we constrain the pinned vortex path to comprise $m$ steps of average $y$-distance $\lambda$, so $y = m\lambda$ is the vortex length in the $y$-direction, and $m$ steps of length $z_i$ in the $z$-direction, where the distribution of lengths $z_i$ is $p(z_i)$. The total $z$ displacement is then given by $z = \sum z_i$, with a distribution $p(z)$. The principle of maximum entropy can be applied and we hypothesize entropy maximizing distributions of $p(z)$ and hence $g(\theta)$, leading to different results for $J_c(\theta)$ through a transformation of random variables from $(y, z)$ to $\theta$. If the distribution $p(z)$ is Gaussian, we obtain
\begin{equation}
J_c(\theta) = \frac{J_0}{\sqrt{2\pi}\Gamma\sin^2\theta}\exp\left(-\frac{1}{2\Gamma^2\tan^2\theta}\right),
\label{eq:gaussian}
\end{equation}
while if $p(z)$ is Lorentzian, we obtain 
\begin{equation}
J_c(\theta) = \frac{1}{\pi}\frac{J_0\Gamma}{\cos^2\theta + \Gamma^2\sin^2\theta}.
\label{eq:lorentzian}
\end{equation}
\end{subequations}
We term Eqs.~(\ref{eq:gaussian}) and (\ref{eq:lorentzian}) angular Gaussian and angular Lorentzian functions, respectively. In Eq.~(\ref{eq:gaussian}), $\Gamma = \sigma/\sqrt{m}\lambda$, where $m\sigma^2$ is the variance of the Gaussian, while in Eq.~(\ref{eq:lorentzian}), $\Gamma = \gamma/\lambda$, where $m\gamma$ is the scale factor of the Lorentzian. Note that Eq.~(\ref{eq:lorentzian}) incorporates the special case $\Gamma = 1$ for which it reduces to the uniform function of Eq.~(\ref{eq:uniform}).

To explain this further, the choice of a Gaussian distribution for $p(z)$ indicates that there exists a population of orthogonal defects that broaden the peak with the only added constraints that $p(z)$ has a mean and a variance, i.e.\ that the lengths of pinned vortex segments in the $z$-direction form a normal distribution. If multiple defect populations exist spanning different length scales, we may achieve a `heavy tailed' statistics; that is, no convergence to a Gaussian and the distribution becomes Lorentzian. In reality this will always be a truncated Lorentzian due to the finite sample size, thus sidestepping any mathematical qualms regarding the use of the Lorentzian distribution. The Lorentzian is a scale independent distribution; consequently the number of steps, $m$, drops out of Eq.\ (\ref{eq:lorentzian}). We note in passing that the vortex paths have a fractal character in this case. We have plotted the entropy for the uniform and angular functions in Appendix A. Some further details on the extrema of Eq.~(\ref{eq:gaussian}) are listed in Appendix B.

This model naturally allows vortex interactions with multiple defect populations to be taken into account. For interactions with multiple Gaussian populations we have $\Gamma^2 = \Gamma_1^2 + \Gamma_2^2 + \ldots$, while for multiple Lorentzians, $\Gamma = \Gamma_1 + \Gamma_2 + \ldots$. These results are derived in Appendix C. Whether a particular peak is due to a unique population of defects or a compound population is a question of interpretation that needs to be addressed through the experimental evidence.

As a method of inference, the principle of maximum entropy further allows us to make the following observations. If we can describe $J_c(\theta)$ using Eqs.~(\ref{eq:VPMequations}) then we have found the correct constraints for the problem and no better description of $J_c(\theta)$ is possible unless we can find weaker constraints and higher entropy functions to fit the data, which is highly unlikely. If $J_c(\theta)$ cannot be described by these equations then further constraints on the microstates exist which have not been taken into account. The model thus provides a direct test of the high-field electron mass anisotropy scaling hypothesis, Eq.~(\ref{eq:massanisotropyscaling}), since the model as presented takes no account of electron mass anisotropy.

\section{\label{sec:results}Results and discussion}

In this section we fit $J_c(\theta)$ data for a wide variety of defect-engineered samples, both LTS and HTS, to combinations of Eqs.~(\ref{eq:VPMequations}). In the vortex path model each equation corresponds to one set of constraints for one set of microstates. There is no expectation that a single equation enumerates all the possible microstates and so fitting involves choosing what, in the field of probability, is termed a mixture distribution. For example, with both $ab$-plane and $c$-axis correlated defects present, both a so-called ``$ab$-plane'' peak and a ``$c$-axis'' peak in $J_c$ are possible. Thus, the combination of two orthogonal defect populations can lead to a mixture of two (or possibly more) distributions. The equations above have been written for a peak centered on the $ab$-plane; for a $c$-axis peak the $\theta$ coordinate will be modified accordingly. Likewise, although not explicitly considered here, it is perfectly possible to account for inclined defect structures, for example due to off-axis irradiation, vicinal substrates, inclined substrate deposition or ion-beam assisted deposition, within the framework of the vortex path model

The fitting has been done ``by eye'' with the assistance of a non-linear least squares fitting algorithm. In some data sets there exist asymmetries or drift which we have not attempted to correct. A formal statistical best fit procedure is possible but such a development is not warranted for the interpretation we are attempting here. More importantly we are attempting to interpret these distributions in terms of the underlying defect structure which contributes to each set of microstates or pinned vortex paths. This interpretation is of necessity tentative due to our limited knowledge of both the microstructure of the samples and the effectiveness of different elements of that microstructure in pinning.

\subsection{\label{sec:Nb}Nanostructured Nb films}

We begin by presenting data obtained on Nb thin films nanostructured to incorporate finely-tuned arrays of vertical pores. The sample preparation details and $J_c(B)$ data for these samples were published elsewhere \cite{Dinner2011}. The samples have a structure of nano-engineered pores etched vertically through the 36 nm thickness of the film with an average pore diameter of 75 nm and an inter-pore spacing of about 140 nm. The $J_c$ values for different samples are variable, but for the best samples approach the depairing current density $J_d$ at self-field. The incorporation of the nano-pores increases $J_c$ by up to a factor of 50 over the virgin films, so clearly the pores contribute significant pinning.

\begin{figure}
\includegraphics{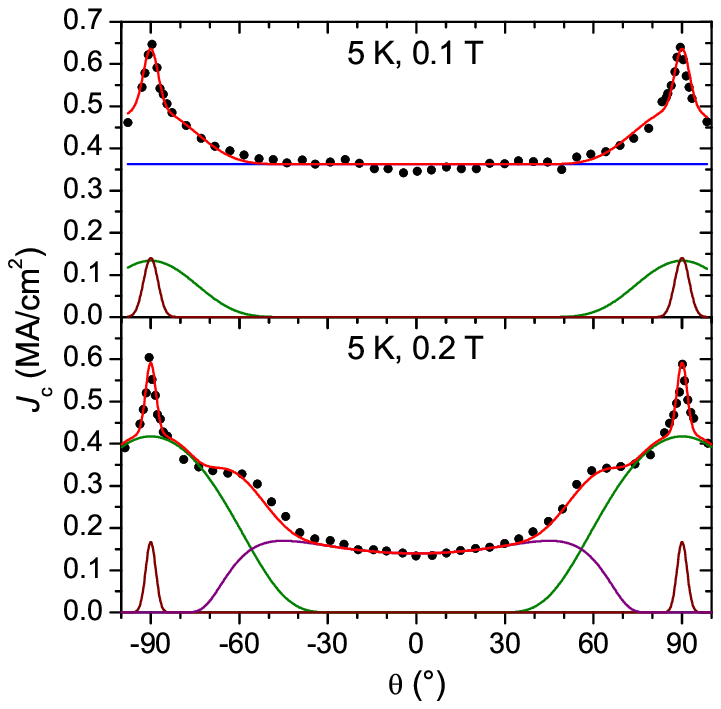}
\captionof{figure}{\label{fig:Nb}$J_c(\theta)$ at 5 K for a Nb thin film incorporating an array of vertical columnar pores: experiment ($\bullet$), full fit (\textcolor{red}{\protect\rule[0.5ex]{0.4cm}{1.5pt}}), fit components (\textcolor{blue}{\protect\rule[0.5ex]{0.4cm}{1.5pt}},\textcolor{purple}{\protect\rule[0.5ex]{0.4cm}{1.5pt}},\textcolor{brown}{\protect\rule[0.5ex]{0.4cm}{1.5pt}},\textcolor{green}{\protect\rule[0.5ex]{0.4cm}{1.5pt}}). The fitting parameters for these data are summarized in Table \ref{tab:Nb}.}
\end{figure}

\begin{figure}
\captionof{table}{\label{tab:Nb}Parameters of fit components in Fig.~\ref{fig:Nb}.}
\begin{ruledtabular}
\begin{tabular}{r@{\extracolsep{1ex}}l@{\extracolsep{\fill}}cdd}
& Angular function & & \multicolumn{1}{c}{\textrm{0.1 T}} & \multicolumn{1}{c}{\textrm{0.2 T}}\\
\colrule
& Uniform (\textcolor{blue}{\rule[0.5ex]{0.4cm}{1.5pt}}) & $J_0$ & 1.140 & -\\
\multirow{2}{*}{OOP} & \multirow{2}{*}{Gaussian (\textcolor{purple}{\rule[0.5ex]{0.4cm}{1.5pt}})} & $J_0$ & - & 0.351\\
& & $\Gamma$ & - & 1.000\\
\multirow{2}{*}{IP} & \multirow{2}{*}{Gaussian (\textcolor{brown}{\rule[0.5ex]{0.4cm}{1.5pt}})} & $J_0$ & 0.015 & 0.013\\
& & $\Gamma$ & 0.043 & 0.030\\
\multirow{2}{*}{IP} & \multirow{2}{*}{Gaussian (\textcolor{green}{\rule[0.5ex]{0.4cm}{1.5pt}})} & $J_0$ & 0.083 & 0.426\\
& & $\Gamma$ & 0.249 & 0.400\\
\end{tabular}
\end{ruledtabular}
\end{figure}

Measurements of $J_c(\theta)$ on one of these samples are shown in Fig.~\ref{fig:Nb}. We note that this data is qualitatively similar to much of the literature data obtained on YBCO and observe that, following the approach of \cite{CivaleJLTP2004,*CivaleAPL2004}, such data would be interpreted as comprising three distinct angular regimes: (i) a sharp in-plane peak in the region of $\pm90^\circ$, attributed to staircase vortices predominantly following the intrinsic pinning $ab$-planes; (ii) a broader underlying in-plane peak attributed to mass anisotropy effects; and (iii) an enhanced $J_c$ near the out-of-plane direction at $0^\circ$ (in this case manifesting itself as a flat angular dependence over an extended angular range), attributed to correlated out-of-plane pinning.

Evidently, in the case of this data on Nb, interpretations (i) and (ii) are manifestly incorrect, and yet the features remain. Applying the vortex path model to this data, we also identify three components. However, our attribution of the origin of these components is somewhat different. In the vortex path model, each component is identified as resulting from the statistical combination of a dominant pinning population, giving correlated pinning aligned with the centre of the peak, and a pinning species providing a pinning force orthogonal to this direction, that broadens the peak. Thus, the two components (brown and green) centered on $\pm90^\circ$ represent two distinct combinations of dominant in-plane pinning with a pinning population orthogonal to this (i.e.\ out-of-plane). Interestingly the two components have similar peak height. We are not certain as to the origin of this effect but it may be due to the frequency of interaction of the vortex path with the orthogonal defects being similar. The two defect populations operate over different length scales (reflected in the widths of the peaks). Since this particular system is extremely simple, clean and well-defined, the only likely sources of correlated pinning are the surface interfaces (the film thickness being comparable to the coherence length) and the patterned vertical pores. We therefore identify the broader peak as arising from the interaction of the surface pinning with large-scale vertical defects comprising the pore structures themselves, while the narrow peak is the result of the interaction of the surface pinning with the (small-scale) surface roughness caused by the presence of the pores. It is then clear why these two interactions should occur with the same frequency.

The third component in the low-field data appears in the form of a constant background. This is the uniform function. A better understanding of the origin of this component comes from considering the behavior under a larger applied field, where it is seen that this component narrows into a rather broad angular Gaussian. The angular peak width of this component is similar to that of the broad in-plane peak, and so we attribute this third component to the dominant out-of-plane pinning facilitated by the pores, broadened by the interaction with the surface pinning. The scale factor is somewhat larger because instead of being limited by the length of the pore (i.e.\ the film thickness), it is now determined by the distance the vortex is able to travel in the in-plane direction while maintaining the macroscopic field orientation. Since this pinning mechanism is effective out to angles of about $\pm60^\circ$, this distance is approximately $\tan 60^\circ = \sqrt{3}$ times the film thickness. At the lower field the weaker vortex-vortex interactions remove the statistical constraints on the majority of vortex paths formed between the pores and the surface pinning, thus resulting in the large uniform contribution.

It is to be emphasized from these results that the effect of the introduction of perfectly strong, perfectly correlated out-of-plane pinning defects is not necessarily the creation of a strong out-of-plane peak in the critical current. The absence of such a peak, as here, cannot therefore be taken as evidence of the absence of correlated pinning. Here, the existence of such pinning centers has a determining effect on the critical current, but it does not manifest itself as a simple out-of-plane peak. Instead it contributes at all angles, through both the broadening of the in-plane peak and the addition of a broad and almost flat component centered on the out-of-plane direction. This is of critical importance for the interpretation of other data, for example in similar experiments to these recently conducted on YBCO \cite{Bagarinao2012} which also lack an out-of-plane peak in $J_c(\theta)$. Likewise, `shoulders' in the angular $J_c$ dependence at an intermediate angle between the in-plane and out-of-plane directions are a natural result of the interplay of in-plane and out-of-plane correlated pinning, and not a signature of some unorthodox type of pinning at an arbitrary angle. As we shall see, such shoulders are an increasingly prominent feature of pinning-optimized samples, and one which the electron mass anisotropy scaling approach is entirely unable to address.

\subsection{\label{sec:PLD}PLD YBCO films}

\begin{figure}
\includegraphics{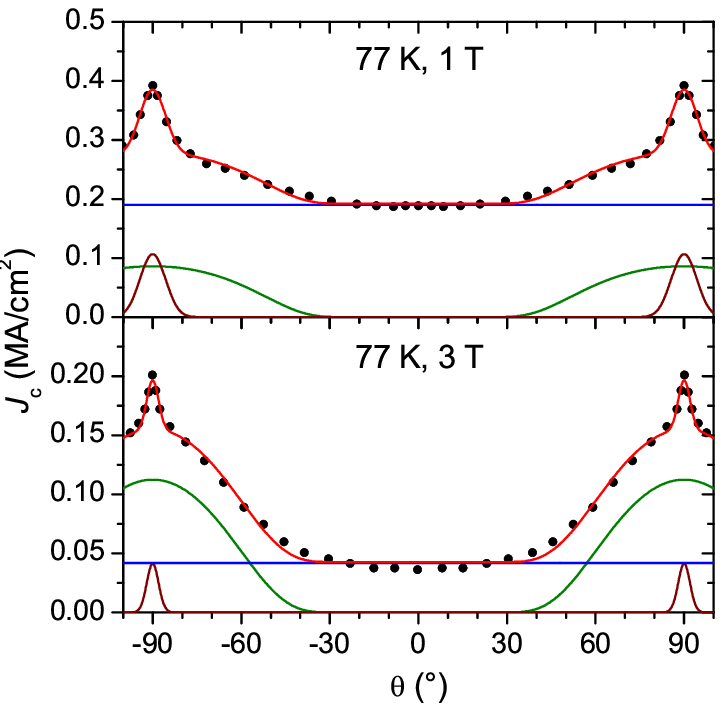}
\captionof{figure}{\label{fig:PLD}$J_c(\theta)$ at 77 K for a PLD YBCO film (data from \cite{Ercolano2011}): experiment ($\bullet$), full fit (\textcolor{red}{\protect\rule[0.5ex]{0.4cm}{1.5pt}}), fit components (\textcolor{blue}{\protect\rule[0.5ex]{0.4cm}{1.5pt}},\textcolor{brown}{\protect\rule[0.5ex]{0.4cm}{1.5pt}},\textcolor{green}{\protect\rule[0.5ex]{0.4cm}{1.5pt}}). The fitting parameters for these data are summarized in Table \ref{tab:PLD}.}
\end{figure}

Turning now to HTS materials, we begin with the simplest data available for an epitaxial 300 nm thick YBCO film, without nanostructural modifications, prepared by pulsed laser deposition (PLD) on a single crystal SrTiO$_3$ substrate \cite{Ercolano2011}, and we see that all the features just described are again present (Fig.~\ref{fig:PLD}). There is no need to invoke anything beyond the known microstructure of the sample to explain all of the observed features. In this thicker film, rather than surface pinning, effective in-plane pinning is provided by the intrinsic pinning due to the planar crystal structure of the sample, with non-superconducting planes separated by roughly the coherence length. Significant out-of-plane pinning is provided by the array of well-known structural defects resulting from the columnar growth mode of PLD samples that are the underlying cause of the high $J_c$ in thin films \cite{Wimbush2009}.

\begin{figure}
\captionof{table}{\label{tab:PLD}Parameters of fit components in Fig.~\ref{fig:PLD}.}
\begin{ruledtabular}
\begin{tabular}{r@{\extracolsep{1ex}}l@{\extracolsep{\fill}}cdd}
& Angular function & & \multicolumn{1}{c}{\textrm{1 T}} & \multicolumn{1}{c}{\textrm{3 T}}\\
\colrule
& Uniform (\textcolor{blue}{\rule[0.5ex]{0.4cm}{1.5pt}}) & $J_0$ & 0.600 & 0.132\\
\multirow{2}{*}{$ab$} & \multirow{2}{*}{Gaussian (\textcolor{brown}{\rule[0.5ex]{0.4cm}{1.5pt}})} & $J_0$ & 0.020 & 0.004\\
& & $\Gamma$ & 0.075 & 0.035\\
\multirow{2}{*}{$ab$} & \multirow{2}{*}{Gaussian (\textcolor{green}{\rule[0.5ex]{0.4cm}{1.5pt}})} & $J_0$ & 0.103 & 0.113\\
& & $\Gamma$ & 0.473 & 0.396\\
\end{tabular}
\end{ruledtabular}
\end{figure}

At 1 T, we observe three contributions to the pinning profile: two $ab$-centered angular Gaussians and a uniform component. This indicates two distinct $c$-axis defect populations, each broadening the $ab$-peak by a different amount. The peak heights of the two Gaussians again coincide, suggesting that the populations of orthogonal defects are correlated with each other as was the case in the Nb films. We hypothesize that the explanation is the same: the population of longer $c$-axis defects (e.g.\ threading dislocations which penetrate the full film thickness) also creates small scale disorder (e.g.\ surface roughness) which vortices lying close to the $ab$-plane orientation see as a defect population available to provide the small amount of $c$-axis pinning required to broaden their peak. If we label the scale factors for these individual populations as $\Gamma_1$ and $\Gamma_2$ then the broad peak is most probably of scale factor $\Gamma = (\Gamma_1^2 + \Gamma_2^2)^{1/2}$, i.e.\ we observe the contribution of pinning paths that interact with both species. Where the two distributions are strongly disparate, as here, the influence of the small scale disorder on the broader peak becomes negligible. Note that the $c$-axis defects present must of necessity also be contributing to the uniform component as this is the only component with a non-zero magnitude at $\theta = 0$. All other defects in the sample also contribute to the uniform component.

At 3 T, the peak magnitudes of the narrow $ab$-peak and the uniform component are the same. The broad $ab$-peak is now of greater magnitude than these components. This is the same trend as was observed for Nb. This suggests that the only effective $c$-axis pinning remaining at this field is now provided by the through-thickness dislocations that match the frequency of the defects broadening the narrow $ab$-peak. These defects remain effective in broadening the $ab$-peak as at 1 T. However, the matching field of these defects is such that they are no longer effective pins in their own right at 3 T, and therefore the uniform component is diminished in magnitude.

\subsection{\label{sec:MOD}MOD YBCO tapes}

The real power of the vortex path model, however, is revealed in its application to technologically-relevant HTS samples. In materials presently being developed for application, pinning effects dominate at all angles, necessitating consideration of the effects of combining multiple different defect populations. It is in these samples that the vortex path model comes into its own, explaining every aspect of the observed data in a coherent, self-consistent manner.

\begin{figure}
\includegraphics{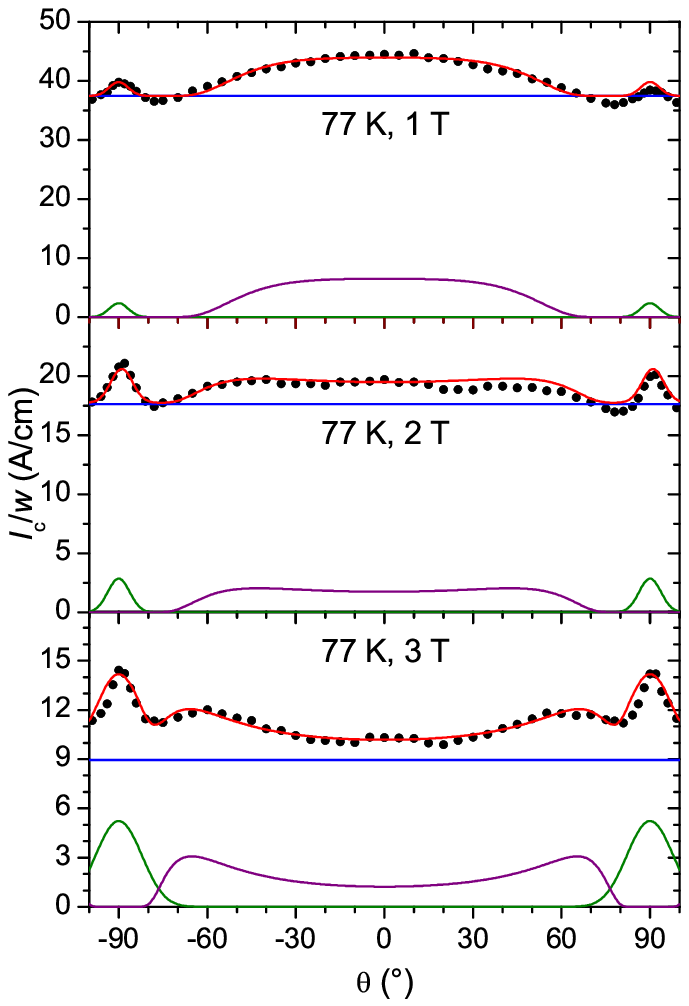}
\captionof{figure}{\label{fig:MOD}$I_c(\theta)$ at 77 K for MOD YBCO tape: experiment ($\bullet$), full fit (\textcolor{red}{\protect\rule[0.5ex]{0.4cm}{1.5pt}}), fit components (\textcolor{blue}{\protect\rule[0.5ex]{0.4cm}{1.5pt}},\textcolor{purple}{\protect\rule[0.5ex]{0.4cm}{1.5pt}},\textcolor{green}{\protect\rule[0.5ex]{0.4cm}{1.5pt}}). The fitting parameters for these data are summarized in Table \ref{tab:MOD}. Where there exists significant asymmetry in the data set, the first half of the data has been fitted.}
\end{figure}

\begin{figure}
\captionof{table}{\label{tab:MOD}Parameters of fit components in Fig.~\ref{fig:MOD}.}
\begin{ruledtabular}
\begin{tabular}{r@{\extracolsep{1ex}}l@{\extracolsep{\fill}}cddd}
& Angular function & & \multicolumn{1}{c}{\textrm{1 T}} & \multicolumn{1}{c}{\textrm{2 T}} & \multicolumn{1}{c}{\textrm{3 T}}\\
\colrule
& Uniform (\textcolor{blue}{\rule[0.5ex]{0.4cm}{1.5pt}}) & $J_0$ & 117.8 & 55.45 & 28.15\\
\multirow{2}{*}{$c$} & \multirow{2}{*}{Gaussian (\textcolor{purple}{\rule[0.5ex]{0.4cm}{1.5pt}})} & $J_0$ & 0.331 & 0.461 & 1.755\\
& & $\Gamma$ & 0.056 & 0.065 & 0.133\\
\multirow{2}{*}{$ab$} & \multirow{2}{*}{Gaussian (\textcolor{green}{\rule[0.5ex]{0.4cm}{1.5pt}})} & $J_0$ & 10.83 & 4.21 & 5.19\\
& & $\Gamma$ & 0.664 & 0.962 & 1.69\\
\end{tabular}
\end{ruledtabular}
\end{figure}

Angular $I_c$ data for a production sample of American Superconductor Amperium\texttrademark tape with a superconducting YBCO layer fabricated by metal-organic deposition (MOD) are shown in Fig.~\ref{fig:MOD}. The major contributions to $I_c(\theta)$ can be fitted with three angular functions at all fields. A well-defined $ab$-plane peak increases in magnitude as the applied field increases. This is paired with an extremely broad $c$-axis peak that broadens further under increasing field, resulting in the characteristic shoulder shape. Note that these shoulders are in approximately the same position as for the LTS film. The gradual broadening of the $c$-axis peak is consistent with the observation that the $ab$-plane pinning is increasing in strength, and will therefore contribute more to the broadening of the $c$-axis peak. Because the fit comprises just two angular Gaussian functions centered on the orthogonal directions, this indicates that there exists a single population of $ab$-plane defects and a single population of $c$-axis defects which are the dominant contributors to the angle-dependent pinning. The two peaks arise from the combination of these same two pinning populations as outlined previously.

The sample additionally has a significant uniform component at all fields. We have seen in the Nb data how this can arise from correlated defects. In these films it may also include a substantial contribution from point defects. An assembly of point defects may pin a single vortex or they may act in combination with correlated defects. To emphasize the importance of this interpretation, which differs from the standard view, we note that in the data measured at 1 T and 2 T there is a small angular region around $\pm75^\circ$ where neither of the angular Gaussian functions has any magnitude. It is not credible to suppose that in precisely this region the correlated defects present in the sample provide no contribution to $I_c$, while contributing significantly at all other angles. As evidence to support this view, we have seen in the PLD sample that the $ab$-peak has a width extending over this region, while in this sample at 3 T we can again see directly the contribution of correlated defects in this region. However, we also know that in these pinning-engineered samples, the incorporation of nanoparticles --- non-correlated defects --- has increased the overall $I_c$. Therefore both nanoparticles and correlated defects must be contributing to the uniform component.

These samples differ from previously described MOD samples \cite{Long2005} in having much stronger $c$-axis pinning. Presumably this is due to twin planes or twin plane intersections, which are the only known high density $c$-axis defects to occur naturally in MOD samples. Again, the vortex path model enables us to relate the pinning profile observed in $I_c(\theta)$ to microstructural features of the sample that can be investigated experimentally.

The fitting obtained for samples on technical substrates using angular Gaussians is seldom as good as for angular Lorentzians, as seen throughout this paper and in \cite{Long2008}. This is because the samples are also disordered in their grain orientation, with an out-of-plane mosaic spread around $5^\circ$. With the Lorentzian components, this actually helps to produce better fits as it generates the same result as having a collection of slightly differing length scales for the same defect population across different grains. As explained in \cite{Long2008}, a Lorentzian is equivalent to a mixture of Gaussians and the disorder helps to create a spread of Gaussians. Conversely, when we have a Gaussian, the grain misorientation is working against a good fit. A Gaussian requires a single scale parameter but we don't have one because defects in differently oriented grains project onto our measurement axes differently. On this basis, it would be appropriate to smooth the fits over an angular range approximating the mosaic spread in the sample. Further counting against a good Gaussian fitting, we will never truly have a single defect population giving all the correlated $J_c(\theta)$ --- there are bound to be minor defect populations also contributing. This is most likely the case in this sample where in the 2 T and 3 T data a second $c$-axis peak can just about be resolved. Nonetheless, even without accounting for these finer details, we obtain a high level of accuracy with very few components.

\subsection{\label{sec:NbO}YBCO with Ba$_2$YNbO$_6$ additions}

\begin{figure}
\includegraphics{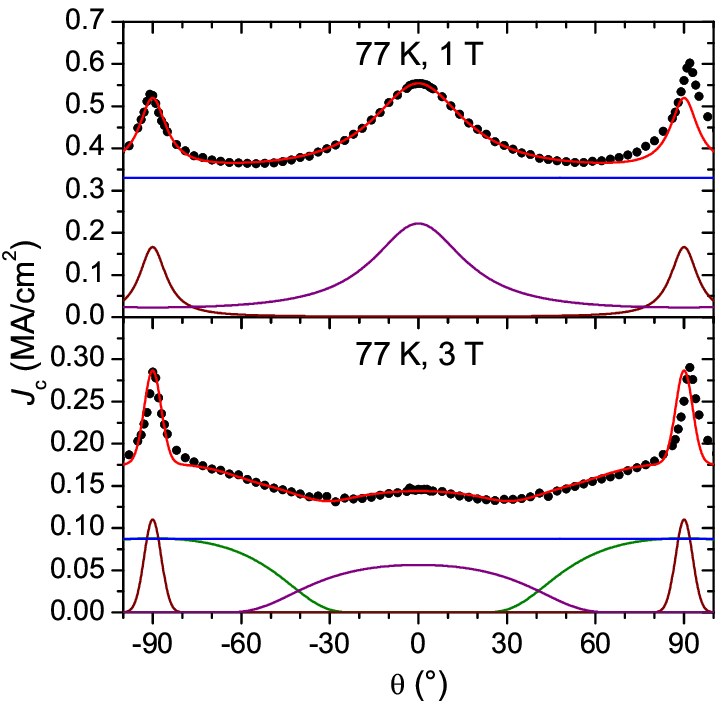}
\captionof{figure}{\label{fig:NbO}$J_c(\theta)$ at 77 K for a PLD YBCO film with Ba$_2$YNbO$_6$ additions (data from \cite{Ercolano2011}): experiment ($\bullet$), full fit (\textcolor{red}{\protect\rule[0.5ex]{0.4cm}{1.5pt}}), fit components (\textcolor{blue}{\protect\rule[0.5ex]{0.4cm}{1.5pt}},\textcolor{purple}{\protect\rule[0.5ex]{0.4cm}{1.5pt}},\textcolor{brown}{\protect\rule[0.5ex]{0.4cm}{1.5pt}},\textcolor{green}{\protect\rule[0.5ex]{0.4cm}{1.5pt}}). The fitting parameters for these data are summarized in Table \ref{tab:NbO}. Where there exists significant asymmetry in the data set, the first half of the data has been fitted.}
\end{figure}

We now turn to defect-engineered samples in the true meaning of the term; that is to say samples with incorporated defects designed to influence the pinning profile in particular ways. The first of these we will consider is the introduction of self-assembled $c$-axis aligned nanocolumns of second-phase material intended to provide strong, effective flux pinning in the $c$ direction where $J_c$ is usually lowest. In the present case, this has been achieved through the addition of the double-perovskite Ba$_2$YNbO$_6$ to PLD YBCO films \cite{Ercolano2010}. This material is incorporated into the YBCO matrix as $\sim$100 nm long, 10-15 nm diameter rods aligned with the $c$-axis of the YBCO and has enabled a two to three times enhancement in $J_c(B{\parallel}c)$ compared with pure YBCO films. The data shown in Fig.~\ref{fig:NbO} are for a 300 nm thick film comprising 5 mol\% Ba$_2$YNbO$_6$, resulting in nanocolumns having an average lateral spacing of around 40 nm.

\begin{figure}
\captionof{table}{\label{tab:NbO}Parameters of fit components in Fig.~\ref{fig:NbO}.}
\begin{ruledtabular}
\begin{tabular}{r@{\extracolsep{1ex}}l@{\extracolsep{\fill}}cdd}
& Angular function & & \multicolumn{1}{c}{\textrm{1 T}} & \multicolumn{1}{c}{\textrm{3 T}}\\
\colrule
& Uniform (\textcolor{blue}{\rule[0.5ex]{0.4cm}{1.5pt}}) & $J_0$ & 1.040 & 0.273\\
\multirow{2}{*}{$c$} & \multirow{2}{*}{Lorentzian / Gaussian (\textcolor{purple}{\rule[0.5ex]{0.4cm}{1.5pt}})} & $J_0$ & \textrm{L }0.223 & \textrm{G }0.075\\
& & $\Gamma$ & \textrm{L }1.012 & \textrm{G }0.524\\
\multirow{2}{*}{$ab$} & \multirow{2}{*}{Lorentzian / Gaussian (\textcolor{brown}{\rule[0.5ex]{0.4cm}{1.5pt}})} & $J_0$ & \textrm{L }0.050 & \textrm{G }0.138\\
& & $\Gamma$ & \textrm{L }0.302 & \textrm{G }0.050\\
\multirow{2}{*}{$ab$} & \multirow{2}{*}{Gaussian (\textcolor{green}{\rule[0.5ex]{0.4cm}{1.5pt}})} & $J_0$ & - & 0.125\\
& & $\Gamma$ & - & 0.570\\
\end{tabular}
\end{ruledtabular}
\end{figure}

At low field (1 T), we once again require three components to fit the data: a uniform component and two Lorentzian peaks. The presence of Lorentzian rather than Gaussian peaks may be caused by additional lattice disorder introduced into the YBCO by the nanocolumns due to their large lattice mismatch and consequent straining of the matrix. Significant distortion of the YBCO lattice has been observed in samples of this composition. Additionally, splaying of the nanorods about the $c$-axis means the direction of their pinning is not well-defined, but instead encompasses a distribution of its own. Consequently, instead of a single defect population yielding a single scale parameter there are multiple Gaussians present which have merged to produce a Lorentzian lineshape. In contrast to the pure YBCO films, here the two peaks are not both $ab$-plane peaks, but rather one of them is a strong $c$-axis peak as a result of the highly effective nanocolumnar pins. In fact, at this field, these pins are so effective that the magnitude of the $c$-axis peak is greater than that of the $ab$-plane peak. This in turn explains the absence of the second $ab$-plane peak in this data: the strength of the $c$-axis pinning is such that it is completely dominant in its interaction with the available $ab$-pinning. This results in the strong uniform component ranging across the angles usually occupied by the broad $ab$-plane peak.

At higher field (3 T), the nanocolumnar pinning begins to lose its strength, a common phenomenon in samples utilizing present flux pinning strategies. This is because the overall density and therefore matching field is relatively low for this defect population. Clearly the pinning centers are still present, but they lose in dominance to the $ab$-plane pinning, and the broad peak associated with their interaction with this pinning source is recovered. The response at this field is extremely similar to the pure YBCO PLD films, with the addition of a small $c$-axis peak due to the residual effect of the additional engineered $c$-axis pinning. The higher field thus again helps to distinguish the different pinning contributions. The change in peak shape from Lorentzian to Gaussian likely results from vortex-vortex interactions at high field acting to constrain the vortices, diminishing both the effectiveness of the pinning and reducing the effect of splay. We also see the previously noted phenomenon of the two $ab$-plane peaks coinciding in magnitude, and additionally here the uniform component also coincides. The uniform component is about twice the magnitude seen in the pure YBCO, reflecting the continuing contribution of the engineered $c$-axis pinning to the uniform component, not just to the $c$-axis peak.

\subsection{\label{sec:TaO}YBCO with Gd$_3$TaO$_7$ additions}

In the attempt to progress beyond BaZrO$_3$ as an effective second-phase pinning inclusion for YBCO, and to increase the effectiveness of this type of artificial pinning center, the pyrochlore rare earth tantalate \emph{R}$_3$TaO$_7$ material was successfully introduced, resulting in significant $J_c$ enhancement compared to BaZrO$_3$-containing samples across the angular range \cite{Harrington2009}. Due to its close lattice match to YBCO, in contrast to BaZrO$_3$ and Ba$_2$YNbO$_6$, \emph{R}$_3$TaO$_7$ addition readily produces extremely fine and highly linear through-thickness nanorods without splaying, and with no poisoning effect reducing the critical temperature. By a simple variation in deposition rate, the nanostructure of the \emph{R}$_3$TaO$_7$ additions can be controlled, either freezing the additions into a fine dispersion of nanoparticles of $\sim$5 nm diameter (at high deposition rate), or allowing time for diffusion-mediated growth into extended nanorods of 5-10 nm diameter and 10-15 nm spacing (at low deposition rate). This provides a unique `lever' offering unprecedented control over the nature of the flux pinning in samples incorporating this additive \cite{Harrington2010}.

\begin{figure*}
\includegraphics{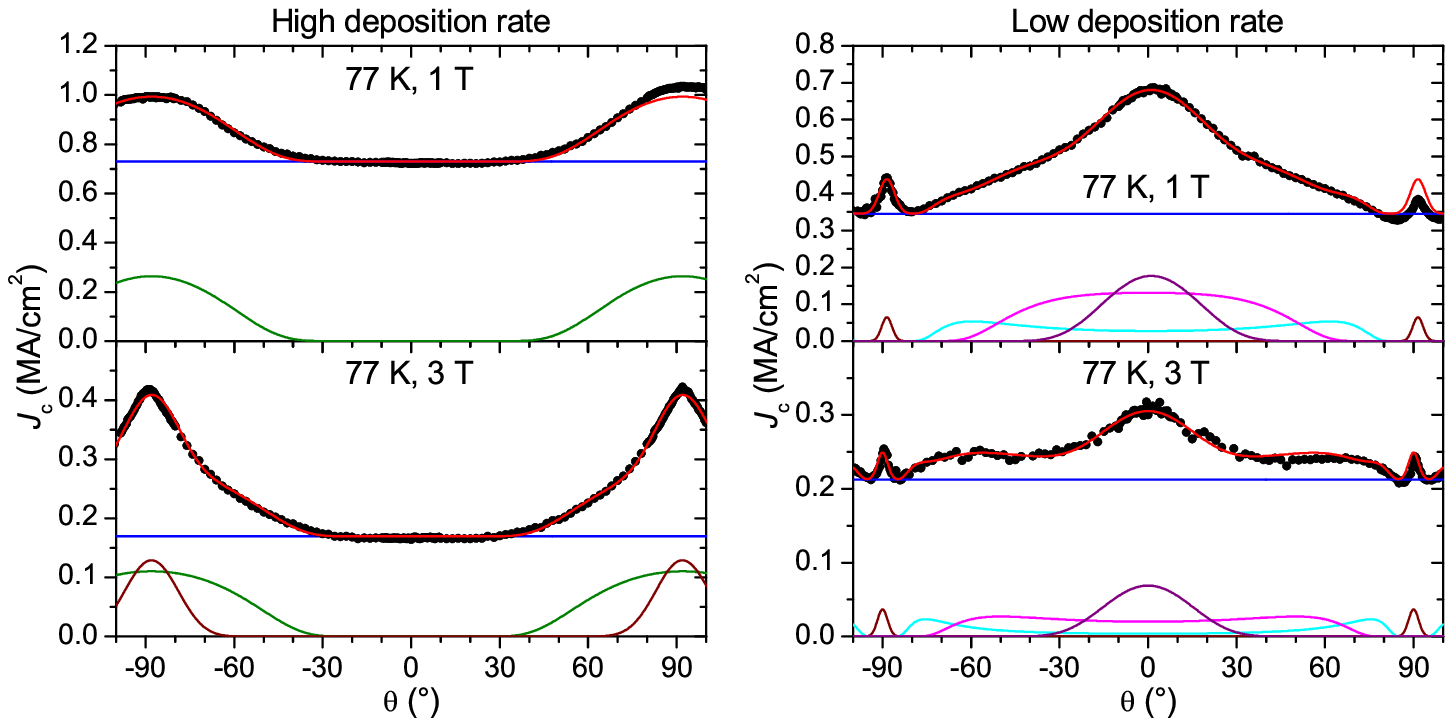}
\captionof{figure}{\label{fig:TaO}$J_c(\theta)$ at 77 K for PLD YBCO films with Gd$_3$TaO$_7$ additions, deposited at different rates to induce nanoparticle or nanorod formation (data from \cite{Harrington2010}): experiment ($\bullet$), full fit (\textcolor{red}{\protect\rule[0.5ex]{0.4cm}{1.5pt}}), fit components (\textcolor{blue}{\protect\rule[0.5ex]{0.4cm}{1.5pt}},\textcolor{purple}{\protect\rule[0.5ex]{0.4cm}{1.5pt}},\textcolor{pink}{\protect\rule[0.5ex]{0.4cm}{1.5pt}},\textcolor{cyan}{\protect\rule[0.5ex]{0.4cm}{1.5pt}},\textcolor{brown}{\protect\rule[0.5ex]{0.4cm}{1.5pt}},\textcolor{green}{\protect\rule[0.5ex]{0.4cm}{1.5pt}}). The fitting parameters for these data are summarized in Table \ref{tab:TaO}.}
\end{figure*}

\begin{figure}
\captionof{table}{\label{tab:TaO}Parameters of fit components in Fig.~\ref{fig:TaO}.}
\begin{ruledtabular}
\begin{tabular}{r@{\extracolsep{1ex}}l@{\extracolsep{\fill}}ccdddd}
& \multirow{2}{*}{Angular function} & & \multicolumn{2}{c}{\textrm{High rate}} & \multicolumn{2}{c}{\textrm{Low rate}}\\
& & & \multicolumn{1}{c}{\textrm{1 T}} & \multicolumn{1}{c}{\textrm{3 T}} & \multicolumn{1}{c}{\textrm{1 T}} & \multicolumn{1}{c}{\textrm{3 T}}\\
\colrule
& Uniform (\textcolor{blue}{\rule[0.5ex]{0.4cm}{1.5pt}}) & $J_0$ & 2.293 & 0.534 & 1.084 & 0.669\\
\multirow{2}{*}{$c$} & \multirow{2}{*}{Gaussian (\textcolor{purple}{\rule[0.5ex]{0.4cm}{1.5pt}})} & $J_0$ & - & - & 0.117 & 0.043\\
& & $\Gamma$ & - & - & 0.263 & 0.244\\
\multirow{2}{*}{$c$} & \multirow{2}{*}{Gaussian (\textcolor{pink}{\rule[0.5ex]{0.4cm}{1.5pt}})} & $J_0$ & - & - & 0.207 & 0.055\\
& & $\Gamma$ & - & - & 0.630 & 1.100\\
\multirow{2}{*}{$c$} & \multirow{2}{*}{Gaussian (\textcolor{cyan}{\rule[0.5ex]{0.4cm}{1.5pt}})} & $J_0$ & - & - & 0.099 & 0.027\\
& & $\Gamma$ & - & - & 1.444 & 2.816\\
\multirow{2}{*}{$ab$} & \multirow{2}{*}{Gaussian (\textcolor{brown}{\rule[0.5ex]{0.4cm}{1.5pt}})} & $J_0$ & - & 0.050 & 0.011 & 0.003\\
& & $\Gamma$ & - & 0.154 & 0.045 & 0.030\\
\multirow{2}{*}{$ab$} & \multirow{2}{*}{Gaussian (\textcolor{green}{\rule[0.5ex]{0.4cm}{1.5pt}})} & $J_0$ & 0.253 & 0.128 & - & -\\
& & $\Gamma$ & 0.384 & 0.465 & - & -\\
\end{tabular}
\end{ruledtabular}
\end{figure}

The data shown in Fig.~\ref{fig:TaO} for 500 nm thick YBCO films prepared by PLD with 1.5 mol\% Gd$_3$TaO$_7$ additions exemplify the pinning control that can be achieved with this additive, where the anisotropy is seen to vary from broad $ab$-peak in the case of a high-rate deposition through to strong $c$-axis peak for low-rate deposition. Supported by TEM imaging, the simple additive model of pinning used in \cite{Harrington2010} to describe this data provides an adequate explanation of the primary features, that is to say the existence of a broad, flat profile with residual $ab$-peaks in the case of the isotropic nanoparticle pinning produced at high growth rate giving way to a strong $c$-axis peak in the presence of the nanocolumns formed at low growth rate. However, application of the vortex path model to these data allows us to decompose the behavior further.

In the case of the sample deposited at high rate, we see the simplest behavior observed to date --- just two components: one uniform and one broad $ab$-peak. Similar to the pure PLD YBCO film, we attribute the broad $ab$-peak to dominant intrinsic $ab$-plane pinning broadened by the type of through-thickness defects characteristic of PLD. The effect of the nanoparticle inclusions is to strongly enhance the uniform component with the result that the sharp $ab$-peak previously observed in films of this type is obscured. This is an example of similar-strength pinning centers adding non-linearly to generate the total pinning force, and evidences the failure of a simple additive pinning model. As the applied field is increased, the effectiveness of the nanoparticulate pinning centers diminishes, leading to a reduction in the uniform component, and the sharp $ab$-plane peak is again distinguished.

Conversely, in the case of the sample deposited at low rate, the most prominent feature is the $c$-axis peak resulting from the nanorods. The vortex path model represents this as three angular Gaussians of increasing width, all centered on the $c$-axis. Here, due to the extreme linearity and crystalline perfection of the nanorods, no tendency towards a Lorentzian lineshape is observed, and distinct broadening defect populations can be resolved. Three components of differing width imply three distinct populations of broadening defects available to interact with the nanorods. Based on the limited microstructural data available on these samples, we speculate that, in common with the MOD samples, strong $ab$-plane pinning is one of the broadening populations, while another broadening defect may be specific $ab$-plane defects such as stacking faults. These tentative assignments can be verified or refuted through experiments designed to control the density of specific defects within the sample and correlate this with $J_c(\theta)$. As the field is increased, these three components become better resolved, broadening and developing the characteristic shoulders also seen in the MOD samples as the intrinsic $ab$-plane pinning becomes relatively more effective than the artificial $c$-axis pinning.

The advantage of the vortex path model in analyzing this data is that we can see how the basic components of correlated and uncorrelated defects are combining to produce the overall $J_c(\theta)$ behavior. That particular peaks are narrow or broad does not mean that whole pinning populations are suddenly appearing or disappearing but rather that the relative strengths and the statistics of how these populations combine is being altered.

\subsection{\label{sec:NbOTaO}YBCO with Ba$_2$YNbO$_6$ + Gd$_3$TaO$_7$ additions}

\begin{figure*}
\includegraphics{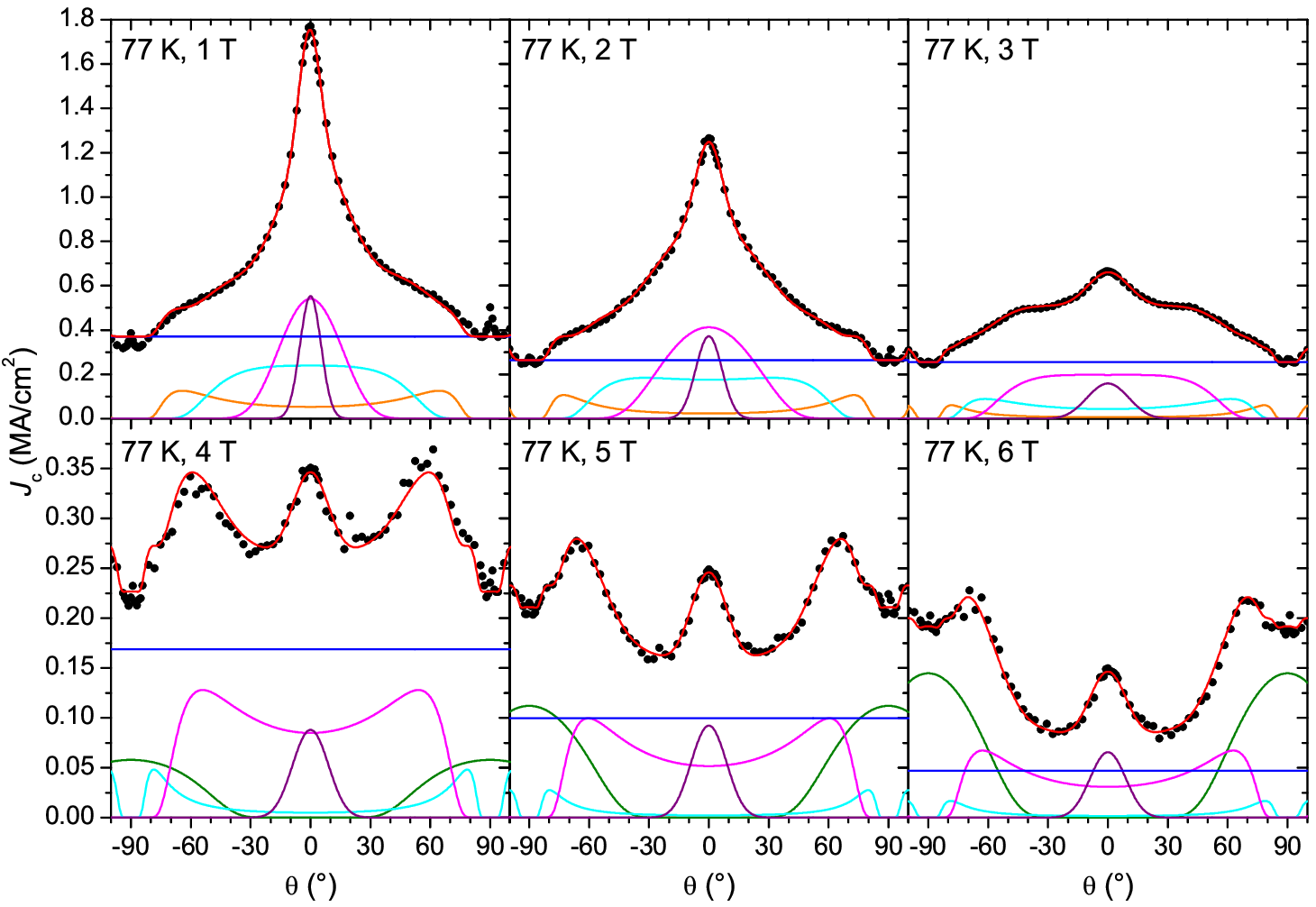}
\captionof{figure}{\label{fig:NbOTaO}$J_c(\theta)$ at 77 K for a PLD YBCO film with Ba$_2$YNbO$_6$ + Gd$_3$TaO$_7$ additions (data from \cite{Ercolano2011}; 2 T data unpublished data from the same sample): experiment ($\bullet$), full fit (\textcolor{red}{\protect\rule[0.5ex]{0.4cm}{1.5pt}}), fit components (\textcolor{blue}{\protect\rule[0.5ex]{0.4cm}{1.5pt}},\textcolor{purple}{\protect\rule[0.5ex]{0.4cm}{1.5pt}},\textcolor{pink}{\protect\rule[0.5ex]{0.4cm}{1.5pt}},\textcolor{cyan}{\protect\rule[0.5ex]{0.4cm}{1.5pt}},\textcolor{orange}{\protect\rule[0.5ex]{0.4cm}{1.5pt}},\textcolor{green}{\protect\rule[0.5ex]{0.4cm}{1.5pt}}). The fitting parameters for these data are summarized in Table \ref{tab:NbOTaO}. The small $ab$-peak has been ignored in the fitting.}
\end{figure*}

Recently, the importance to pinning optimization of achieving segmented nanocolumnar pins rather than continuous nanocolumns was suggested \cite{Chen2009} to mitigate the effects of thermally activated depinning. One method by which this has been achieved is through the simultaneous incorporation of the two types of pinning additive already considered, Ba$_2$YNbO$_6$ and Gd$_3$TaO$_7$. What forms in this case is a complex composite of YBCO, Ba$_2$Y(Nb,Ta)O$_6$ segmented nanorods, in-plane Y$_2$O$_3$ platelets and nanoparticles of the YBa$_2$Cu$_4$O$_8$ superconducting phase, where the Gd partially substitutes for Y in all four phases and the Ta and Nb content goes entirely into forming nanorods \cite{Ercolano2011}. The $J_c(\theta)$ dependence as a function of the applied field for a 300 nm thick sample of this composition prepared by PLD is shown in Fig.~\ref{fig:NbOTaO}.
 
\begin{figure*}
\captionof{table}{\label{tab:NbOTaO}Parameters of fit components in Fig.~\ref{fig:NbOTaO}.}
\begin{ruledtabular}
\begin{tabular}{r@{\extracolsep{1ex}}l@{\extracolsep{\fill}}cdddddd}
& Angular function & & \multicolumn{1}{c}{\textrm{1 T}} & \multicolumn{1}{c}{\textrm{2 T}} & \multicolumn{1}{c}{\textrm{3 T}} & \multicolumn{1}{c}{\textrm{4 T}} & \multicolumn{1}{c}{\textrm{5 T}} & \multicolumn{1}{c}{\textrm{6 T}}\\
\colrule
& Uniform (\textcolor{blue}{\rule[0.5ex]{0.4cm}{1.5pt}}) & $J_0$ & 1.166 & 0.829 & 0.801 & 0.531 & 0.311 & 0.148\\
\multirow{2}{*}{$c$} & \multirow{2}{*}{Gaussian (\textcolor{purple}{\rule[0.5ex]{0.4cm}{1.5pt}})} & $J_0$ & 0.125 & 0.105 & 0.070 & 0.033 & 0.035 & 0.025\\
& & $\Gamma$ & 0.091 & 0.113 & 0.176 & 0.153 & 0.151 & 0.152\\
\multirow{2}{*}{$c$} & \multirow{2}{*}{Gaussian (\textcolor{pink}{\rule[0.5ex]{0.4cm}{1.5pt}})} & $J_0$ & 0.348 & 0.384 & 0.356 & 0.256 & 0.185 & 0.118\\
& & $\Gamma$ & 0.257 & 0.371 & 0.722 & 1.21 & 1.43 & 1.55\\
\multirow{2}{*}{$c$} & \multirow{2}{*}{Gaussian (\textcolor{cyan}{\rule[0.5ex]{0.4cm}{1.5pt}})} & $J_0$ & 0.404 & 0.368 & 0.163 & 0.045 & 0.023 & 0.015\\
& & $\Gamma$ & 0.672 & 0.831 & 1.49 & 3.56 & 4.00 & 3.64\\
\multirow{2}{*}{$c$} & \multirow{2}{*}{Gaussian (\textcolor{orange}{\rule[0.5ex]{0.4cm}{1.5pt}})} & $J_0$ & 0.218 & 0.140 & 0.055 & - & - & -\\
& & $\Gamma$ & 1.63 & 2.38 & 3.50 & - & - & -\\
\multirow{2}{*}{$ab$} & \multirow{2}{*}{Gaussian (\textcolor{green}{\rule[0.5ex]{0.4cm}{1.5pt}})} & $J_0$ & - & - & - & 0.070 & 0.118 & 0.145\\
& & $\Gamma$ & - & - & - & 0.491 & 0.422 & 0.400\\
\end{tabular}
\end{ruledtabular}
\end{figure*}

From the low field data, the most striking feature is the large sharp $c$-axis peak, far stronger and better defined than we have seen in any other data. Analysis in terms of the vortex path model begins to give us some idea of why this is --- the coincidence of no less than four components centered on the $c$ direction combine to give this peak. Its sharpness is caused by the combination of two narrow components of similar magnitude, while its baseline is boosted by the contribution of two additional broad components, as well as a large uniform component. The effect of this large uniform component is to make the $ab$-peak look small, but once again this is not due to any particular lack of $ab$-pinning in these samples, but rather the strong relative dominance of the $c$-axis pinning. In the same way that strong $ab$-pinning can obscure a $c$-axis peak despite the existence of significant $c$-axis pinning, uncommonly strong $c$-axis pinning as has been achieved here can also obscure the $ab$-peak. The uniform component is similar in magnitude to the nanorod-containing samples of the previous section. 

By analogy with the earlier niobate addition data, we identify the broader of the two sharp $c$-axis peaks as resulting from the segmented nanorods, in interaction with the intrinsic $ab$-plane pinning, observing that it varies with field in the same way, diminishing and broadening out to a relatively flat response by 3 T. This leaves the sharper $c$-axis peak to be identified, and we note that it tracks the broader peak in magnitude. It also remains relatively constant throughout the entire field range, changing little in width or relative magnitude. This suggests two things: firstly that it has a high matching field, i.e.\ it results from a high-density field-independent population of defects, and secondly that its width probably arises from a property such as the angular spread of the nanorods. Hence this is an unusual peak in that it may arise from a single defect species only, interacting with itself through its own disorder.

The broader $c$-axis components then likely arise due to interactions between the segmented nanorods and the in-plane Y$_2$O$_3$ platelets and YBa$_2$Cu$_4$O$_8$ nanoparticles. These unique features are a direct result of the unique microstructure of these samples, with the Y$_2$O$_3$ platelets being of similar dimension to the nanorod segments. It is therefore unsurprising that they greatly extend the angular range over which the nanorods can effectively pin, facilitating pinning of vortices between adjacent nanorod segments at high field angles.

These four $c$-axis components can be traced through to 3 T, by which point the $c$-axis pinning is significantly reduced in strength, as is commonly observed for all artificial pinning centers known to date. The two broad $c$-axis components continue broadening strongly with increasing applied field. To higher fields, the dominance of the $c$-axis pinning is overcome, and familiar features relating to the $ab$-pinning begin to emerge. The resultant combination of these similar strength features yields extremely complex angular variations in $J_c$ that are nonetheless resolvable into a few primitive components that describe the subtlest features of the data extremely well.

At 4 T, an interesting crossover occurs. It is at this field that the strongest nanorod component is broadened to such an extent that it has developed significant shoulders that result in a quite notable response in the $J_c(\theta)$. It is appropriate that such significant microstructural features should continue to give a marked response in the $J_c(\theta)$ even though their contribution to the overall $J_c$ is now diminished. At the same time, the broadest $c$-axis peak vanishes, to be replaced with the usual $ab$-pinning peak that we attribute to intrinsic pinning being broadened by through-thickness defects. The disappearance is assumed to be a consequence of the applied field becoming greater than the matching field of the relevant $c$-axis defect species.

We note a couple of interesting points about the fitting. Again we observe coincidences in the peak heights of multiple components. These are seen to occur not only at the center of the distribution but also at other points where the distributions reach a local extremum. At an extremum $\partial g/\partial \theta = 0$ and the distribution is locally unconstrained. This coincidence may therefore have an interpretation in terms of the maximization of entropy for a mixture distribution which we do not presently understand. The other point we make is to remind the reader that each component has only a single scale parameter. This means that both the shape and the position of the shoulders are highly constrained by the form of the angular Gaussian distribution we have derived. It is difficult to conceive of other primitive functions which could so well reproduce these features of this data.

\section{\label{sec:conclusion}Conclusion}

We have addressed the question of whether the application of the electron mass anisotropy scaling approach to field angular $J_c(\theta)$ data of pinning-optimized samples is theoretically justified, and have concluded that it is not. In experimental support of this determination, we have presented data measured on thin films of the isotropic superconductor Nb that exhibit features commonly attributed to electron mass anisotropy, reinforcing the fact that $J_c$ is a wholly extrinsic material parameter and demanding an alternative explanation that can be related to the microstructure or other extrinsic properties of the sample. We provide this explanation in terms of the vortex path model, and have applied the same model to a wide range of contemporary pinning-engineered samples, from which a number of general results emerge. Firstly, that the introduction of perfectly strong, perfectly correlated out-of-plane pinning defects does not necessarily result in a strong out-of-plane peak in the $J_c(\theta)$, and therefore that the absence of such a peak cannot be taken as evidence of the absence of correlated pinning. Secondly, that pinning from different sources is not a simple additive process, but rather depends on what pinning is already present. In this context, common manifestations of strong out-of-plane correlated pinning are a broadening of the in-plane peak or an almost flat angular variation in $J_c(\theta)$ about the out-of-plane direction, while commonly-observed shoulders in the angular $J_c$ dependence are a result of the interplay between in-plane and out-of-plane correlated pinning of similar strength, and not a signature of exotic pinning at an arbitrary angle. In particular we emphasize that a broad in-plane peak is evidence of strong out-of-plane correlated pinning, and not a reflection of electron mass anisotropy.

Our assignment of defect populations to particular peaks is based on the available knowledge of the sample microstructure. Correct identification of the individual pinning populations requires a correlation between $J_c(\theta)$ and microstructural analysis over a family of samples in which the density of these defect populations is systematically varied and the effect on the different components of $J_c$ quantified. These hypotheses can then be tested by measuring samples in those temperature and field ranges which best distinguish between the different defect populations. It appears from our analysis of available data that measuring the field dependence of the angular profile is very useful for determining the origin of pinning effects. The evolution of the magnitude of a component allows us to determine the matching field for a contributing defect population. The evolution of the shape also correlates with matching field effects, e.g.\ out-of-plane components broaden towards the in-plane direction as the field increases due to the high matching field of the intrinsic pinning.

In the context of the principle of maximum entropy, the success of the vortex path model in describing the data shows that no further constraints beyond those we have imposed on the vortex path statistics are necessary to explain $J_c(\theta)$. Thus we can conclude with confidence that the electronic mass anisotropy plays no role in determining the form of $J_c(\theta)$ for any of the various samples studied. If we find samples which do not immediately fit the vortex path model equations then different constraints are operating in those systems. An obvious example is the need to extend the model to cases where $J_c(\theta)$ has asymmetries due to oblique correlated defect populations. The success of the vortex path model in this instance also suggests that other angular data influenced by the pinning landscape could be better described by Eqs.~(\ref{eq:VPMequations}), or at least other expressions derived from maximum entropy considerations. The Ginzburg-Landau mass anisotropy expression is often applied simply because no alternative angular expressions have been proposed. For example, we can see no reason why the angular dependence of the irreversibility field or the flux flow resistivity should follow an expression involving electronic mass anisotropy.

\begin{acknowledgments}
The authors are grateful to R.\ B.\ Dinner for supplying the Nb thin film samples, and to G.\ Ercolano and S.\ A.\ Harrington for providing digital data sets of their measurements, including some previously unpublished data. We acknowledge helpful discussions with J.\ H\"{a}nisch on the application of the electron mass anisotropy scaling procedure. This work was funded by the New Zealand Ministry of Science and Innovation.
\end{acknowledgments}

\appendix

\section{Entropy of the angular distributions}

\begin{figure}
\includegraphics{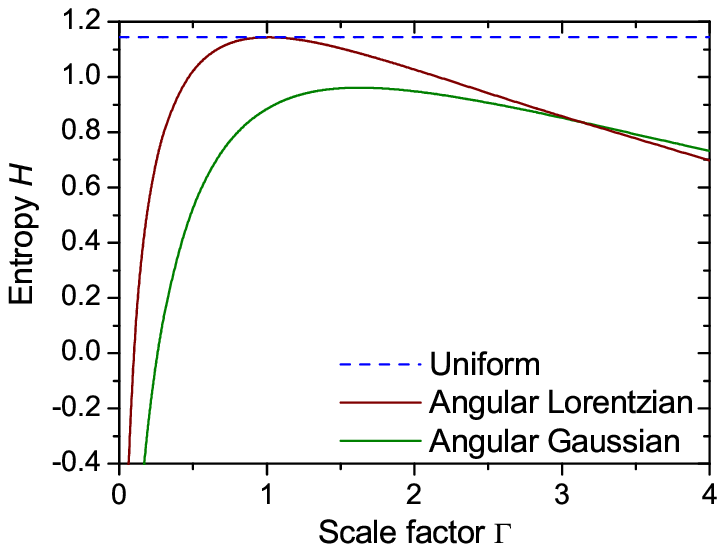}
\addtolength{\belowcaptionskip}{-6mm}
\addtolength{\abovecaptionskip}{-8mm}
\caption{\label{fig:entropy}Entropy of the uniform, angular Lorentzian and angular Gaussian distributions in dependence on their scale factor.}
\end{figure}

The entropy of the normalized uniform, angular Gaussian and angular Lorentzian distributions derived as Eqs.~(\ref{eq:VPMequations}) is calculated from Eq.~(\ref{eq:entropy}). This is done numerically as no analytical solution exists. For discrete distributions $H \ge 0$  but for continuous distributions this restriction is relaxed with the understanding that all entropies are relative quantities. The results are plotted on Fig.~\ref{fig:entropy}. Most striking is the special case whereby the entropy of the angular Lorentzian with scale factor $\Gamma = 1$ equals that of the uniform distribution --- a set of vortices constrained to pin along correlated defects can, in principle, result in a critical current distribution having the same entropy as that resulting from a set of fully unconstrained vortices.

\section{The extrema of Eq.\ (\ref{eq:gaussian})}

Equation (\ref{eq:gaussian}) has extrema at $\theta = 0$ and $\theta = \pm \pi/2$, and for $\Gamma > 1/\sqrt{2}$, at $\theta = \sin^{-1}(1/\sqrt{2}\Gamma)$. The magnitude at $\theta = 0$ is $J_c = 0$, the global minimum. At $\theta = \pm\pi/2$, the magnitude is $J_c = J_0/\Gamma$, which is a maximum for $\Gamma \le 1/\sqrt{2}$ and a local minimum for $\Gamma > 1/\sqrt{2}$. At $\theta = \sin^{-1}(1/\sqrt{2}\Gamma)$, $J_c = 2J_0\Gamma\exp((1/2\Gamma^2)-1)$, a maximum.

\section{Summation of pinning populations}

Suppose we have multiple pinning populations broadening a peak which individually in the vortex path model would give $\Gamma_1 = \sigma_1/(\sqrt{m_1}\lambda_1)$, $\Gamma_2 = \sigma_2/(\sqrt{m_2}\lambda_2)$, etc. A vortex path now interacts with these multiple populations simultaneously. Then we have $y = m_1\lambda_1 = m_2\lambda_2 = \ldots \equiv m\lambda$ where $\lambda$ is now the average spacing between interactions with any $z$-defect from these populations. The scale factors of the Gaussians $p(z)$ combine by the usual convolution rule to give a Gaussian with $m\sigma^2 = m_1\sigma_1^2 + m_2\sigma_2^2 + \ldots$. Then $\Gamma^2 = \sigma^2/(m\lambda^2) = \Gamma_1^2 + \Gamma_2^2 + \ldots$. Likewise, for the addition of Lorentzian components we have $m\gamma = m_1\gamma_1 + m_2\gamma_2 + \ldots$ and $\Gamma = \Gamma_1 + \Gamma_2 + \ldots$. We always fit angular Lorentzians with $\Gamma \le 1$, however there is no reason why $\Gamma > 1$ is forbidden. The equation is still well-behaved and the peak moves to the orthogonal direction. This is simply an alternative description of the statistics of the vortex paths.

\bibliography{apssamp}% Produces the bibliography via BibTeX.

\end{document}